\documentclass{emulateapj}
\usepackage{apjfonts}
\usepackage{graphicx}
\usepackage{xspace}
\usepackage{graphicx}
\usepackage{longtable}
\usepackage{psfig}

\newcommand{\Msun}      {\mbox{$\,M_{\mathord\odot}$}}

\begin{document}

\lefthead{Alternative Explanations for Extreme Supersolar Iron Abundances}
\righthead{Tomsick et al.}

\def\lsim{\mathrel{\lower .85ex\hbox{\rlap{$\sim$}\raise
.95ex\hbox{$<$} }}}
\def\gsim{\mathrel{\lower .80ex\hbox{\rlap{$\sim$}\raise
.90ex\hbox{$>$} }}}

\title{Alternative explanations for extreme supersolar iron abundances inferred 
from the energy spectrum of Cygnus X-1}

\author{John A. Tomsick\altaffilmark{1}, 
Michael L. Parker\altaffilmark{2}, 
Javier A. Garc\'{i}a\altaffilmark{3,4}, 
Kazutaka Yamaoka\altaffilmark{5},
Didier Barret\altaffilmark{6,7},
Jeng-Lun Chiu\altaffilmark{8},
Ma\"{i}ca Clavel\altaffilmark{1,9}, 
Andrew Fabian\altaffilmark{10}, 
Felix F\"{u}rst\altaffilmark{2}, 
Poshak Gandhi\altaffilmark{11},
Victoria Grinberg\altaffilmark{12}, 
Jon M. Miller\altaffilmark{13},
Katja Pottschmidt\altaffilmark{14,15},
Dominic J. Walton\altaffilmark{10}
}

\altaffiltext{1}{Space Sciences Laboratory, 7 Gauss Way, University of California, Berkeley, CA 94720-7450, USA (e-mail: jtomsick@ssl.berkeley.edu)}

\altaffiltext{2}{European Space Agency (ESA), European Space Astronomy Centre (ESAC), E-28691 Villanueva de la Ca$\tilde{\rm n}$ada, Madrid, Spain}

\altaffiltext{3}{Cahill Center for Astronomy and Astrophysics, California Institute of Technology, Pasadena, CA 91125, USA}

\altaffiltext{4}{Remeis Observatory \& ECAP, Universit{\"a}t Erlangen-N{\"u}rnberg, 96049 Bamberg, Germany}

\altaffiltext{5}{Institute for Space-Earth Environmental Research (ISEE) and Division of Particle and Astrophysical Science, Graduate School of Science, Nagoya University, Furo-cho, Chikusa-ku, Nagoya, Aichi 464-8601, Japan}

\altaffiltext{6}{Universit\'{e} de Toulouse; UPS-OMP; IRAP; Toulouse, France}

\altaffiltext{7}{CNRS; Institut de Recherche en Astrophysique et Plan\'{e}tologie; 9 Av. colonel Roche, BP 44346, F-31028 Toulouse cedex 4, France}

\altaffiltext{8}{Institute of Astronomy, National Tsing Hua University, Hsinchu 30013, Taiwan}

\altaffiltext{9}{Universit\'{e} Grenoble Alpes, CNRS, IPAG, F-38000 Grenoble, France}

\altaffiltext{10}{Institute of Astronomy, University of Cambridge, Madingley Road, Cambridge CB3 0HA, UK}

\altaffiltext{11}{Department of Physics and Astronomy, University of Southampton, Southampton SO17 3RT, UK}

\altaffiltext{12}{ESA European Space Research and Technology Centre (ESTEC), Keplerlaan 1, 2201 AZ Noordwijk, The Netherlands}

\altaffiltext{13}{Department of Astronomy, University of Michigan, 500 Church Street, Ann Arbor, MI 48109, USA}

\altaffiltext{14}{CRESST and NASA Goddard Space Flight Center, Astrophysics Science Division, Code 661, Greenbelt, MD 20771, USA}

\altaffiltext{15}{Center for Space Science and Technology, University of Maryland Baltimore County, 1000 Hilltop Circle, Baltimore, MD 21250, USA}

\begin{abstract}

Here we study a 1--200\,keV energy spectrum of the black hole binary Cygnus~X-1 taken 
with {\em NuSTAR} and {\em Suzaku}.  This is the first report of a {\em NuSTAR}
observation of Cyg~X-1 in the intermediate state, and the observation was taken
during the part of the binary orbit where absorption due to the companion's
stellar wind is minimal.  The spectrum includes a multi-temperature thermal disk
component, a cutoff power-law component, and relativistic and non-relativistic 
reflection components.  Our initial fits with publicly available constant density 
reflection models ({\ttfamily relxill} and {\ttfamily reflionx}) lead to extremely 
high iron abundances ($>$9.96 and $10.6^{+1.6}_{-0.9}$ times solar, respectively).
Although supersolar iron abundances have been reported previously for Cyg~X-1, our
measurements are much higher and such variability is almost certainly unphysical.
Using a new version of {\ttfamily reflionx} that we modified to make the electron
density a free parameter, we obtain better fits to the spectrum even with solar
iron abundances.  We report on how the higher density 
($n_{e} = (3.98^{+0.12}_{-0.25})\times 10^{20}$\,cm$^{-3}$) impacts other parameters 
such as the inner radius and inclination of the disk.

\end{abstract}


\section{Introduction}

The bright black hole (BH) high mass X-ray binary Cygnus~X-1 has been key for many of 
the advances in our knowledge of accreting BHs.  A mass measurement of the compact object 
showing that it is well above the maximum mass allowed for a neutron star ($>$3\Msun) 
provided the first evidence that such compact objects exist \citep{gb86}.  The fact 
that BHs enter into different spectral states, such as the soft and hard states, was 
first demonstrated with observations of Cyg~X-1 \citep{tananbaum72}.  Cyg~X-1 is one 
of two sources where it has been possible to show that the compact jet, which is seen 
in the hard state, is an extended radio feature \citep{stirling01}.  Cyg~X-1 is also 
known to have a strong power-law component in its spectrum extending to at least a few
MeV \citep{mcconnell02} with a very high level ($\sim$70\%) of polarization 
\citep{laurent11,jourdain12}, possibly indicating that the emission above 400\,keV is 
dominated by synchrotron emission from the jet.

The properties of Cyg~X-1 are very well-constrained compared to most BH binaries.  It 
has a parallax distance measurement of $1.86^{+0.12}_{-0.11}$\,kpc \citep{reid11}, which 
is consistent with a dust-scattering measurement \citep{xiang11}.  Optical and near-IR 
observations made over its 5.6 day orbit constrain the BH mass to be $14.8\pm 1.0$\Msun~and 
its binary inclination to be $27.1^{\circ}\pm 0.8^{\circ}$ \citep{orosz11}.  The spin of 
the BH has been constrained using two techniques: modeling the thermal continuum 
and the reflection component.  The measurements agree that the spin of the BH is high.
The continuum technique gives a value for the spin ($a_{*} > 0.983$, 3-$\sigma$ limit),
which is consistent with being maximal or very slightly below maximal \citep{gou14},
while the reflection measurements allow for somewhat more modest spin
\citep{duro11,miller12,fabian12,tomsick14,walton16}.  For example, using four
observations, \cite{walton16} find a range of $a_{*}$ values from 0.93 to 0.96.
One reason that the \cite{walton16} values are a little lower than the \cite{gou14}
values is that \cite{gou14} assume that the inner disk is aligned with the
orbital plane.

The \cite{gou14} and \cite{walton16} spin measurements occur when the source is in
the soft state and rely on the assumption that the disk extends to the innermost
stable circular orbit (ISCO).  Measurements of Cyg~X-1 and other systems, such as
LMC~X-3 \citep{steiner10}, provide good evidence that the disk extends to the ISCO
in the soft state.  However, the situation is less clear in the hard and intermediate
states.  For the hard state, it is predicted that thermal conduction of heat from the 
corona will cause the inner disk to evaporate \citep{mlm00}, leading to an optically
thick and geometrically thin disk that is truncated at some significant distance away 
from the black hole.  While there is evidence for truncation at luminosities near
0.1\% of the Eddington limit ($L_{\rm Edd}$) for GX~339--4 \citep{tomsick09c},
measurements of the reflection component in the bright hard state ($\gsim$5\% $L_{\rm Edd}$)
often lead to estimates of inner radii very close to the ISCO for several sources,
including GRS~1739--278 \citep{miller15}.  The luminosity of Cyg~X-1 is typically
0.8--2.8\% $L_{\rm Edd}$, using 0.1--500\,keV unabsorbed fluxes \citep{yamada13b}, and
even with high quality data from the {\em Nuclear Spectroscopic Telescope Array (NuSTAR)}
and {\em Suzaku}, there have been conflicting results concerning whether the disk in
the hard state is truncated or not \citep{parker15,basak17}.  

Given that the binary inclination for Cyg~X-1 has been measured to high precision,
it is also interesting to compare the inner disk inclinations inferred from
reflection measurements to the binary value ($27.1^{\circ}\pm 0.8^{\circ}$).
In the soft state, \cite{tomsick14} found an inner disk inclination $>$40$^{\circ}$,
and \cite{walton16} reported values between $38^{\circ}$ and $41^{\circ}$ for four
different soft state measurements.  Taken at face value, the data suggest a
disk warp of 10--15$^{\circ}$, which has important implications for BH formation
since it is very likely that the BH would need to be formed with its spin
misaligned from the binary plane to produce this warp \citep{bp75,sm94,fragile07}.
Another implication is that the Cyg~X-1 BH would need to be born with its rapid
rotation rate (as described above, $a_{*} = 0.93$--0.96 for a warped disk or
$a_{*} > 0.983$ if the entire disk is aligned with the orbital plane).

Another surprising result from the reflection fits for Cyg~X-1 as well as
GX~339--4 is the measurement of high iron abundances.  For example,
\cite{walton16} find a value of $A_{\rm Fe} = 4.0$--4.3 times solar for
Cyg~X-1, and \cite{parker15} find $A_{\rm Fe} = 4.7\pm 0.1$.  For GX~339-4,
values of $A_{\rm Fe} = 5\pm 1$ and $6.6^{+0.5}_{-0.6}$ have been found
\citep{garcia15,parker16}.  \cite{fuerst15} also obtain
high values of $A_{\rm Fe}$ for GX~339--4, but they find that the
abundances change significantly for different assumptions about the
continuum fitting.  V404~Cyg is another black hole system where reflection
fits have resulted in supersolar abundances of $A_{\rm Fe} = 2$--5 
\citep{walton17}.

While there have been previous reports of {\em NuSTAR} and {\em Suzaku}
observations of Cyg~X-1 in the soft state and the hard state
\citep{tomsick14,parker15,walton16,basak17}, the work in this paper presents
the first analysis of contemporaneous observations of Cyg~X-1 in the
intermediate state with these satellites.  As the name implies, the
intermediate state has luminosities and spectral parameters (such as disk
temperatures and power-law slopes) that are in between the soft and hard
states, and it may provide clues to how sources in general and Cyg~X-1
in particular make transitions between soft and hard states.  The Cyg~X-1
stellar wind can cause significant absorption of the X-ray spectrum,
including distortion of the iron line \citep{tomsick14,walton16}, but the
time of the intermediate state observation was carefully chosen to occur
at the orbital phase when the BH is in front of the companion to minimize
the absorption.  In section 2, we describe the observations and how the
data were analyzed.  Section 3 includes the results of the spectral
fitting, and the results are discussed in Section 4.

\section{Observations and Data Reduction}

\begin{figure*}
\plotone{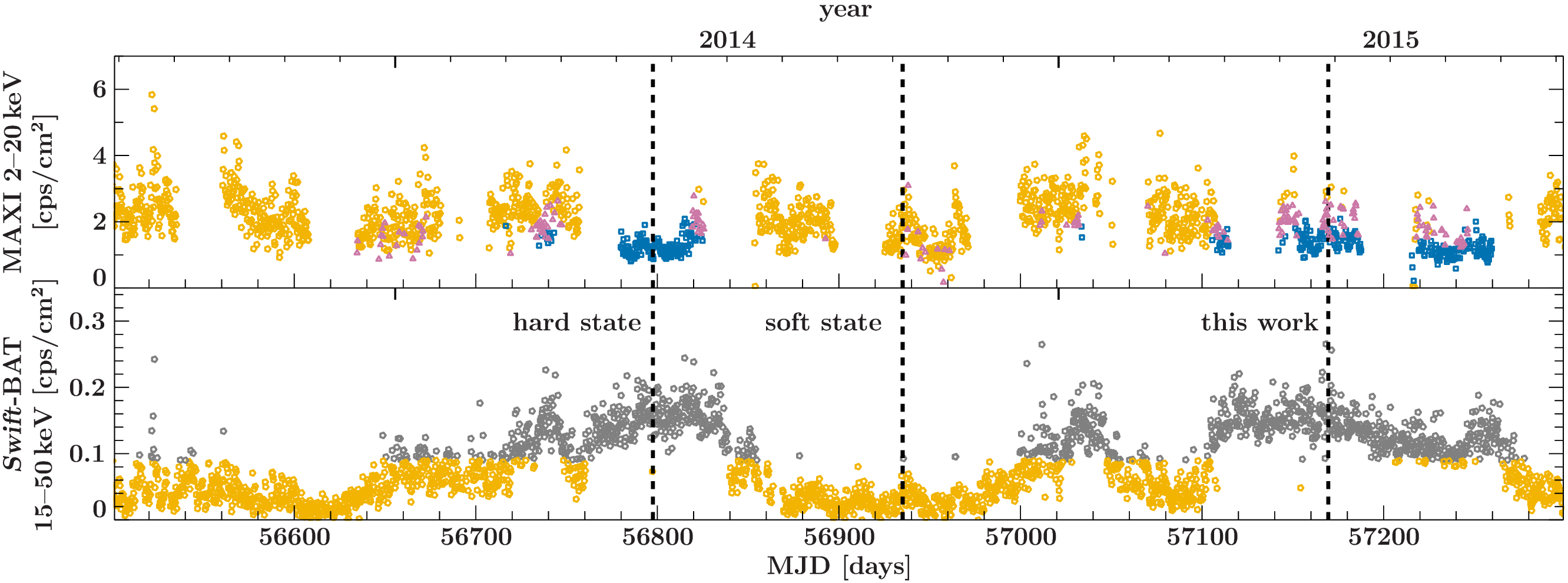}
\caption{\small {\em MAXI} light curve in the 2--20\,keV band for Cyg~X-1 and {\em Swift}/BAT 
light curve in the 15--50\,keV band, both binned to 6 hours. For both instruments, yellow 
circles represent soft states, magenta triangles intermediate states and blue squares hard 
states. For the {\em Swift}/BAT light curve, gray circles indicate either hard or intermediate 
states that cannot be easily distinguished in the BAT energy range. The vertical lines mark 
the times of {\em NuSTAR} observations from this work, hard state \citep{parker15,basak17}
and soft state \citep{walton16}.\label{fig:lc_longterm}}
\end{figure*}

We observed Cyg~X-1 with {\em NuSTAR} \citep{harrison13} and {\em Suzaku}
\citep{mitsuda07} on 2015 May 27--28 (MJD 57,169--57,170).  Light curves measured
by the {\em Monitor of All-sky X-ray Image} \citep[{\em MAXI};][]{matsuoka09}
and the {\em Swift} Burst Alert Telescope \citep[BAT;][]{krimm13} in the 2--20 and
15--50\,keV energy bands, respectively, are shown in Figure~\ref{fig:lc_longterm}.  
They cover the time of the observation that we are focusing on in this work as 
well as the earlier hard state observation reported on in \cite{parker15} and 
\cite{basak17}.  The BAT count rate was similar for the two observations, but 
the 2--20\,keV {\em MAXI} count rate was higher during the 2015 May observation.  
\cite{grinberg13} used Cyg~X-1 data from all-sky monitors, including {\em MAXI}, 
to define count rates for different states.  The weighted average of the count 
rates for {\em MAXI} measurements within 3 days of the 2015 May observation gives 
2--4 and 4--10\,keV rates of $1.032\pm 0.011$ and $0.654\pm 0.006$\,s$^{-1}$, 
respectively, and the ratio of the 4--10 to 2--4\,keV rates is $0.633\pm 0.009$.  
These values indicate that Cyg~X-1 was in the intermediate state during the 
2015 May observation.  The states are indicated in Figure~\ref{fig:lc_longterm}.

Table~\ref{tab:obs} provides information about the {\em NuSTAR} and {\em Suzaku}
observations used in this work, the observation identifier numbers (ObsIDs), the
start and stop times of the observations, and the exposure times obtained.  The
middle of the observation is close to the orbital phase corresponding to the
superior conjunction of the supergiant, and the range of orbital phases covered
is 0.42--0.56.  This was planned in order to minimize absorption due to stellar
wind material.  In the following, we describe the reduction of the data from each
mission in more detail.

\subsection{NuSTAR}

We processed the data for the two {\em NuSTAR} focal plane modules (FPMA and FPMB) 
using HEASOFT v6.21, NUSTARDAS v1.7.0, and version 20170503 of the {\em NuSTAR}
calibration data base (CALDB).  We produced cleaned event files using
{\ttfamily nupipeline} and then light curves and spectra using {\ttfamily nuproducts}.  
For light curves and spectra, we used a circular source extraction region with
a radius of $180^{\prime\prime}$ and a circular background region with a radius
of $90^{\prime\prime}$.  Due to the brightness of Cyg~X-1, the background region was
placed on a corner of the {\em NuSTAR} field of view away from the source.  We
rebinned the spectra with the requirement that the source is detected at the
30-$\sigma$ level or higher in each bin.  

\begin{figure*}
\plotone{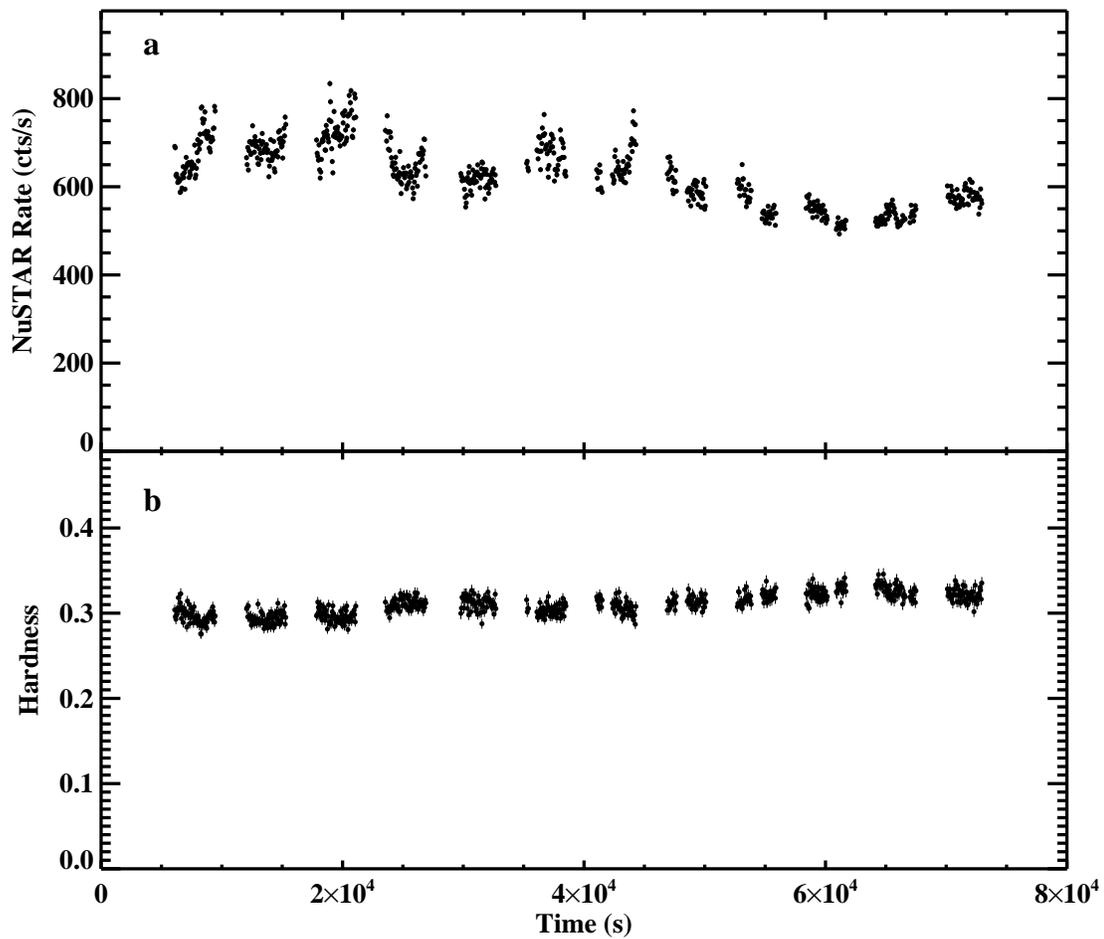}
\caption{\small {\em (a)} {\em NuSTAR} 3--79\,keV light curve with FPMA and FPMB 
rates added together and {\em (b)} {\em NuSTAR} hardness ratio (10--79/3--10\,keV)
for Cyg~X-1.  The time bins are 50\,s in duration.\label{fig:lc2}}
\end{figure*}

The 3--79\,keV light curve (Figure~\ref{fig:lc2}a) shows variability in the FPMA+FPMB
count rate with the lowest 50\,s bin having a rate of 493\,s$^{-1}$ and the highest
having 834\,s$^{-1}$.  However, the hardness (Figure~\ref{fig:lc2}b), which is the
ratio of the 10--79 to the 3--10\,keV rates, shows very little variation, indicating
that the energy spectrum is relatively stable during the observation.  While
{\em NuSTAR} observed the source for $\sim$36\,ks, the actual livetime was
$\sim$20\,ks (Table~\ref{tab:obs}).

\subsection{Suzaku}

For the X-ray Imaging Spectrometers \citep[XIS0, XIS1, and XIS3;][]{koyama07},
we used the 1/4 window mode, which has a CCD readout time of 2\,s.  Also, due to
the brightness of the source, we specified a burst mode to reduce photon pile-up,
causing the CCD to be exposed for 0.3\,s of each 2\,s window.  However, the
detectors were not put in burst mode for the first 19\,ks of the observation,
making this part of the observation is not usable.  The livetimes listed in
Table~\ref{tab:obs} for the XIS units are also affected by filtering out times
when the count rates were high enough to produce telemetry saturation.  It is
this filtering that causes the livetimes to be significantly different for the
individual XIS units.

We reprocessed the XIS data using {\ttfamily aepipeline} and version 20160607
of the CALDB to make new event lists.  We then made a new good time interval
(GTI) file and applied it to the event lists to filter out the first 19\,ks of
the observation.  We used {\ttfamily aeattcor2} to calculate attitude corrections
and applied the corrections with {\ttfamily xiscoord}.  Even with the burst mode,
there is significant photon pile-up.  We calculated the level of pile-up in each
pixel using {\ttfamily pileest}, and made a source extraction region that does
not include the inner part of the point spread function.  We excluded data from
an inner circular region so that all the data that we did include comes from
pixels where the pile-up fraction is less than 4\%.  The other criterion for
a pixel to be included in the source extraction region is that it is within
$4^{\prime}$ of the source position.  For background subtraction, we extracted
a spectrum from rectangular regions that are as far as possible from the source.
We made response matrices for each XIS unit using {\ttfamily xisrmfgen} and
{\ttfamily xissimarfgen}.  XIS0 and XIS3 have very similar responses because
they are both front-illuminated CCDs, and we combined them for spectral fitting.
As in \cite{tomsick14}, we added 2\% systematic uncertainties in quadrature
with the statistical uncertainties.  We rebinned the spectra with the requirement
that the source is detected at the 20-$\sigma$ level or higher in each bin.  

The hard X-ray detector \citep[HXD;][]{takahashi:07a} does not have imaging capabilities
and consists of two kinds of detectors: Silicon PIN diodes covering 12--70\,keV and GSO
scintillators covering 40--600\,keV.  After 2014, {\em Suzaku} operation suffered from
degradation of the power supply system, including the Solar Array Panel (SAP) and
batteries\footnote{See http://heasarc.gsfc.nasa.gov/docs/suzaku/news/battery.html
 and http://heasarc.gsfc.nasa.gov/docs/suzaku/news/power.html}.  For the observations
from 2014 until the end of the mission, it was difficult to use both XIS and HXD in most
cases.  However, we did use both instruments for the {\em Suzaku} observation in 2015 May.
The observation started at 11:23 (UT) on May 27, but the HXD detector parameters were not 
set properly until 17:15 on May 27 (UT) due to the recovery from an ``Under Voltage Control'' 
or UVC event.  Hence, we excluded the data during this period. In addition, to save 
satellite power, HXD was sometimes turned off, leading to a net HXD exposure of 18.6\,ks.  
Cyg~X-1 was observed at the XIS nominal position, i.e., 3.5$^{\prime}$ offset from the HXD 
optical axis.  The HXD deadtime-corrected spectra were obtained in the standard manner 
using standard FTOOLs {\tt hxdpinxbpi} and {\tt hxdgsoxbpi} for PIN and GSO, respectively.  
For both PIN and GSO source spectra, a 1\% systematic error was added for each spectral 
bin \citep[e.g.,][]{torii11}.  The predicted non X-ray background (NXB) was produced from 
files at the public ftp sites\footnote{ftp://legacy.gsfc.nasa.gov/suzaku/data/background/pinnxb\_ver2.2\_tuned/ and ftp://legacy.gsfc.nasa.gov/suzaku/data/background/gsonxb\_ver2.6/}.  
To estimate the systematic error on the GSO background, we compared the data during 
the Earth-occultation period to the model for the background.  The difference between the 
data during occultation and the model lead to an estimate of 5\% systematic error on the GSO
background, and we included this during spectral fitting.  For PIN, we further subtracted 
the cosmic X-ray background (CXB) based on previous observations \citep{gruber99}. 
We used the PIN spectrum from 15--55\,keV except for the 40--45\,keV range due to the
Gd K line structure \citep[see][]{kouzu13}.  For GSO, we used data in the 82--192\,keV
band, with the upper end of the range being set by the statistical quality of the spectrum.
In the spectral fitting, we used the following energy responses: ae\_hxd\_pinxinome11\_20110601.rsp 
for PIN and ae\_hxd\_gsoxinom\_20100524.rsp with an additional correction file 
(ae\_hxd\_gsoxinom\_crab\_20100526.arf) for GSO.

\section{Results}

\begin{figure*}
\plotone{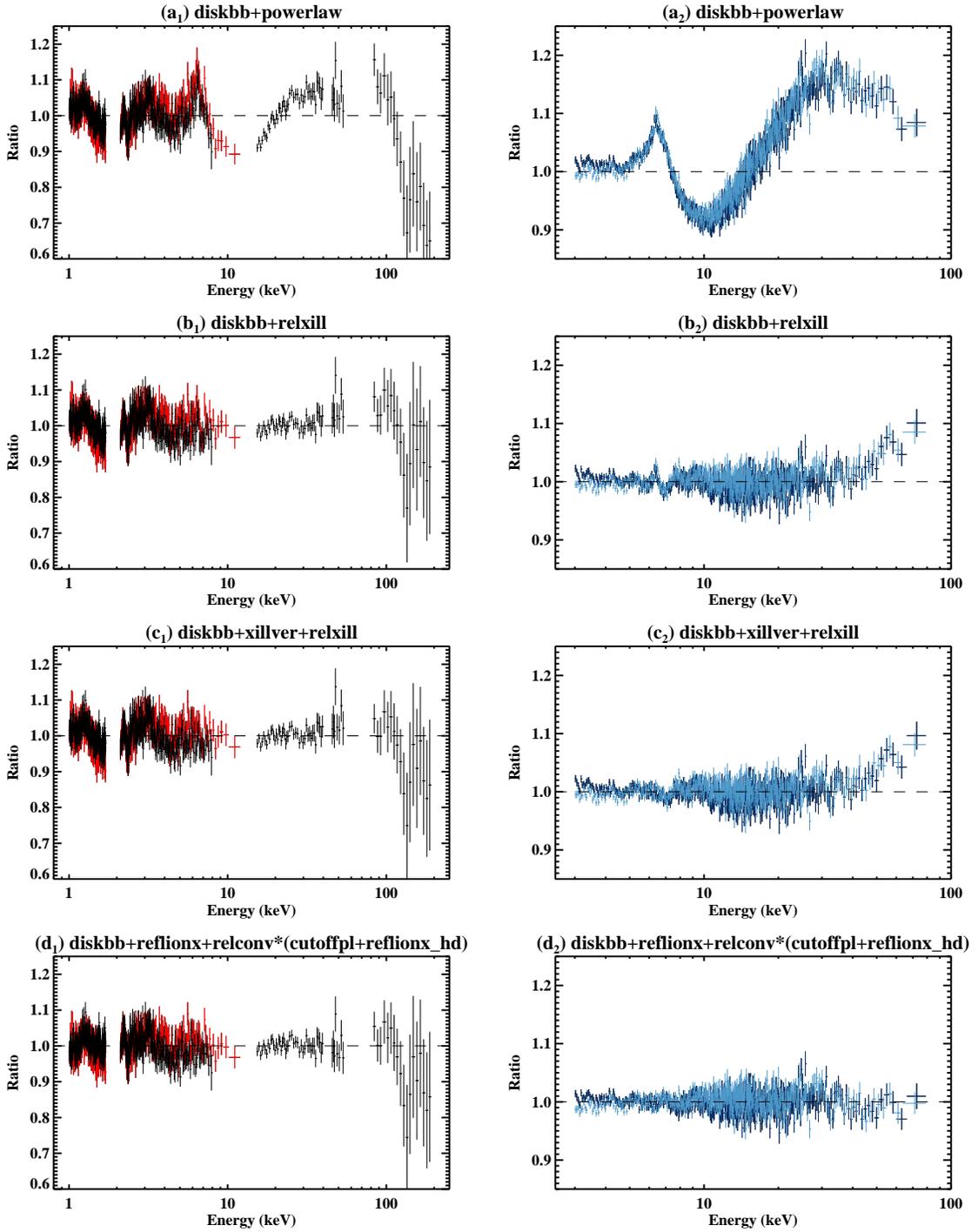}
\caption{\small Data-to-model ratio plots for the four models indicated above each 
plot.  The {\em Suzaku} (left panels) and {\em NuSTAR} (right panels) spectra
were fitted together, but the data-to-model ratios are shown separately for clarity.  
The left panels ($a_{1}$, $b_{1}$, $c_{1}$, and $d_{1}$) show the XIS data 
(red and black), the PIN data, and the GSO data at the highest energies.  The 
right panels ($a_{2}$, $b_{2}$, $c_{2}$, and $d_{2}$) show FPMA (dark blue) and 
FPMB (light blue).  The XIS and {\em NuSTAR} spectra are binned in XSPEC (only
for visual clarity) to levels of 30-$\sigma$ per bin and 50-$\sigma$ per bin,
respectively.\label{fig:ratio}}
\end{figure*}

We initially fitted the {\em NuSTAR}+{\em Suzaku} energy spectrum with a model consisting
of a multi-temperature thermal disk ({\ttfamily diskbb}) component \citep{mitsuda84} and a
power-law.  We also included interstellar absorption using v2.3.2 of the
{\ttfamily tbnew}\footnote{See http://pulsar.sternwarte.uni-erlangen.de/wilms/research/tbabs}
model and a multiplicative constant to allow for differences in normalization between the
six spectra (XIS0+XIS3, XIS1, FPMA, FPMB, PIN, and GSO).  For modeling the absorption,
we used \cite{wam00} abundances and \cite{vern96} cross sections.  A very poor fit is
obtained with $\chi^{2} = 29,415$ for $\nu = 2759$ degrees of freedom (dof).  
Figure~\ref{fig:ratio} shows the data-to-model ratio residuals for this model.  In each case,
the {\em Suzaku} and {\em NuSTAR} spectra were fitted 
simultaneously, but we show two panels per model so that the data from each instrument is 
visible.  The residuals for the {\ttfamily diskbb+powerlaw} model (see 
Figures~\ref{fig:ratio}a$_{1}$ and \ref{fig:ratio}a$_{2}$) show the hallmarks of relativistic Compton 
reflection with positive features at the Fe K$\alpha$ transition energies and in the 
20--100\,keV range.  In the reflection scenario, these features are identified as fluorescence 
of iron in the accretion disk \citep{fabian89} and the ``Compton hump'' \citep{lw88}.  The 
negative residuals in the $\sim$8--15\,keV range are partly due to the surrounding positive 
features and partly caused by a relativistically smeared Fe absorption edge.

Replacing the power-law with a {\ttfamily relxill} component leads to a huge improvement
in the fit (to $\chi^{2}/\nu = 3715/2751$). The {\ttfamily relxill} model includes a
direct power-law with an exponential cutoff as well as the reflection component predicted
when a cutoff power-law is incident on an optically thick accretion disk.  The calculation
of the reflection component is based on the {\ttfamily xillver} model \citep{garcia13,garcia14},
which includes continuum, emission lines, and absorption edges.  This component is then
relativistically smeared, including the effects of the {\ttfamily relline} convolution
calculation \citep{dauser13}.  In this paper, we used v1.0.0 of the {\ttfamily relxill}
model\footnote{See http://www.sternwarte.uni-erlangen.de/$\sim$dauser/research/relxill/}.
Although the residuals shown in Figures~\ref{fig:ratio}b$_{1}$ and \ref{fig:ratio}b$_{2}$ 
are much improved, there is evidence for a narrow iron emission line, which has been seen 
previously \citep{miller02c,walton16}, and is likely due to X-ray irradiation of the 
companion star, the stellar wind material, or the outer disk.

Assuming that the narrow line is caused by irradiation of some cool material in the system,
it is physically reasonable to assume that it is part of a full reflection component,
motivating the addition of a {\ttfamily xillver} component to the model.  A model consisting
of {\ttfamily diskbb+relxill+xillver} significantly improves the quality of the fit to
$\chi^{2}/\nu = 3566/2750$.  Here, the only additional free parameter is the
{\ttfamily xillver} normalization.  We assume that the cool material producing the
{\ttfamily xillver} component is neutral, and we fix the log of the ionization
parameter ($\xi$) to zero.  The other {\ttfamily xillver} parameters are tied to
the {\ttfamily relxill} values.  All the parameters for this fit are shown in
Table~\ref{tab:parameters1}.  While most parameters are free, some are fixed or
restricted to avoid searching physically unreasonable regions of parameter space.
We fixed the interstellar column density, which has been measured many times
\citep[e.g.,][]{tomsick14}, to $N_{\rm H} = 6\times 10^{21}$\,cm$^{-2}$.  Also, we restricted
the spin of the black hole, $a_{*}$, to be $>$0.93 based on previous measurements
\citep{gou14,walton16}.  The {\ttfamily relxill} model assumes that the source emissivity
has a broken power-law dependence on radial distance from the black hole with an index
of $q_{\rm in}$ inside the break radius ($R_{\rm break}$) and $q_{\rm out}$ outside.  We fix
$q_{\rm out}$ to a value of 3.0, and we restrict $q_{\rm in}$ to be greater than $q_{\rm out}$.
Although the GSO data shows that there is a cutoff to the spectrum, the errors on the GSO
points are large compared to {\em NuSTAR}, and this causes problems with constraining
the exponential cutoff energy ($E_{\rm cut}$) and the cross-normalization constant
($C_{\rm GSO}$).  Thus, we fix the value of $E_{\rm cut}$, and the value we use
is 300\,keV in order to allow more direct comparisons to results from the {\rm reflionx}
model \citep{rf05}, which are described below.

While the addition of the {\ttfamily xillver} component improves the fit and
eliminates the residuals near 6.4\,keV, there are still significant residuals
in the {\em NuSTAR} spectrum (see Figures~\ref{fig:ratio}c$_{1}$ and \ref{fig:ratio}c$_{2}$).
Specifically, negative residuals can be seen in the iron edge region near 7\,keV,
and there are positive residuals at the high energy end.  Although the high energy
{\em NuSTAR} residuals could be reduced somewhat (but not completely eliminated) by
using a larger value of $E_{\rm cut}$, this would worsen the fit to the GSO data.
Looking at the model parameters in Table~\ref{tab:parameters1}, another concern 
is the very high iron abundance of nearly 10 times solar abundances 
($A_{\rm Fe} > 9.96$)\footnote{In this paper, we quote limits and errors at 90\% 
confidence unless otherwise indicated.}.  Although supersolar iron abundances have 
been seen previously when fitting the reflection spectra of Cyg~X-1 in the hard 
and soft states \citep{tomsick14,parker15,walton16,basak17}, those values ($A_{\rm Fe}$ 
between 2 and 5) were significantly lower than the value of 10 that we are seeing 
for the intermediate state spectrum.  We also fit the spectrum with the direct
component being a Comptonization continuum ({\ttfamily relxillCp}) instead of
a cutoff power-law, but the best fit value for $A_{\rm Fe}$ is still at its
maximum value of 10.

Although some X-ray binaries may have supersolar abundances, it is difficult 
to explain why $A_{\rm Fe}$ would be variable for Cyg~X-1.  For GRS~1915+105, 
short-term ($\sim$10--100\,s) variations in $A_{\rm Fe}$ have been claimed
\citep{zoghbi16}, and it was suggested that these may be caused by levitation
of the iron atoms by radiation pressure \citep{reynolds12,zoghbi16}.  However, 
the luminosity of Cyg~X-1 ($\sim$1--2\% of the Eddington limit) is much lower
than for GRS~1915+105, indicating that the effect should be weaker for Cyg~X-1.
Also, even for GRS~1915+105, the periods where the iron abundances were found
to be high ($A_{\rm Fe}\sim 3$) lasted only for tens of seconds, and the duty
cycle for supersolar abundances was small.  Thus, the fact that our reflection
fits suggest such a high iron abundance requires us to consider what other
physics is missing from the reflection models that could cause an artificial
increase in $A_{\rm Fe}$.  

One assumption that is made in the constant density reflection models, such
as {\ttfamily relxill}, is that the actual density does not have a large
effect on the predicted spectrum.  The rationale for this is that although
the radiation will penetrate more deeply (in physical length units) into a 
low-density disk than a high-density disk, the penetration is not different 
in terms of optical depth.  However, \cite{garcia16} have recently looked 
at possible density effects, including the fact that free-free absorption
and heating of the disk both depend quadratically on density.  As density
increases, the rise in free-free absorption leads to an increase in
temperature, causing extra thermal emission from the outer layers of the
disk \citep{garcia16}.  Thus, a black hole binary that has a disk with
$n_{e}\sim 10^{20}$\,cm$^{-3}$ will have a soft X-ray excess compared to a
disk with $n_{e} = 10^{15}$\,cm$^{-3}$, which, motivated by values expected
for Active Galactic Nuclei (AGN), is the density assumed for {\ttfamily relxill}.

In order to investigate the effects of higher disk density, we switch from 
{\ttfamily relxill} to {\ttfamily reflionx} \citep{rf05} because a version
of {\ttfamily reflionx} that extends the range of $n_{e}$ to above
$10^{20}$\,cm$^{-3}$ has been produced \citep{rf07}, and we have the capability
to make additional customized models using the same computer code.  While
a high-density version of {\ttfamily relxill} (called {\ttfamily relxillD})
is available, the maximum density considered in {\ttfamily relxillD} is only
$10^{19}$\,cm$^{-3}$.  Prior to switching to the high-density {\ttfamily reflionx}
model, we re-fitted the Cyg~X-1 spectrum with the standard {\ttfamily reflionx}
model \citep{rf05}, which assumes a density of $10^{15}$\,cm$^{-3}$.  The model
is {\ttfamily diskbb+reflionx+relconv*(cutoffpl+reflionx)}, where the direct 
component, {\ttfamily cutoffpl}, and the reflection component, {\ttfamily reflionx},
are convolved with {\ttfamily relconv}, which is the same relativistic convolution
model that is part of {\ttfamily relxill}.  In addition to using {\ttfamily reflionx}
for the relativistic reflection component, we also replaced {\ttfamily xillver}
with {\ttfamily reflionx}.  We fix $\log{\xi}$ to 1.0, which is the minimum value
for {\ttfamily reflionx}, and verified that this component has a narrow emission
line at 6.4\,keV, implying that it is still a valid way to model reflection
from cool material.  An advantage of switching to the {\ttfamily reflionx}
model is that the range of $A_{\rm Fe}$ values goes to 20, allowing us to see
how much this parameter increases beyond 10.

\begin{figure*}
\plotone{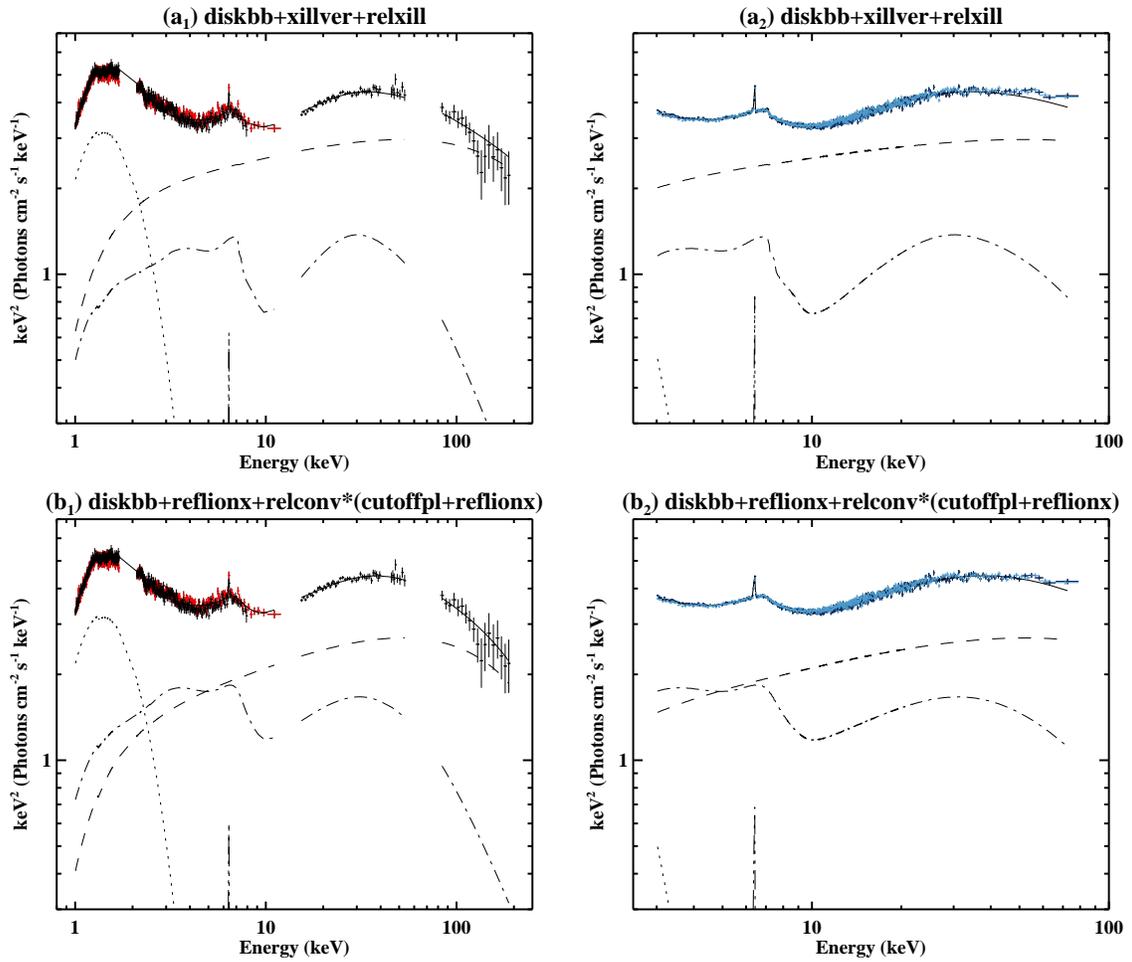}
\caption{\small Unfolded {\em Suzaku} (left panels) and {\em NuSTAR} (right panels)
spectra fitted with two models with the iron abundance parameter free.  In all 
panels, the components are {\ttfamily diskbb} ({\em dotted line}), 
{\ttfamily cutoffpl} ({\em dashed line}), and reflection component ({\em dash-dotted 
line}).  Although a full {\ttfamily reflionx} model is also included, only the narrow
iron line is visible.  The {\em solid line} is the total of all the components.  The
residuals for panels $a_{1}$ and $a_{2}$ are shown in Figure~\ref{fig:ratio}c$_{1}$
and \ref{fig:ratio}c$_{2}$.  The XIS and {\em NuSTAR} spectra are binned in XSPEC
(only for visual clarity) to levels of 30-$\sigma$ per bin and 50-$\sigma$ per bin,
respectively.\label{fig:spectra}}
\end{figure*}

Table~\ref{tab:parameters1} shows that the parameter values are very
similar for the {\ttfamily relxill}-based and {\ttfamily reflionx}-based
models.  In particular, the {\ttfamily reflionx} model still requires a
very high iron abundance, which is well-constrained at $A_{\rm Fe} = 10.6^{+1.6}_{-0.9}$.
However, some differences are also apparent, and these can be seen in
Figure~\ref{fig:spectra}.  With {\ttfamily reflionx}, the reflection component
is somewhat stronger relative to the direct component.  The extra emission from
the reflection component in the soft X-ray band in the model allows the power-law
index to harden slightly (from $\Gamma = 1.826\pm 0.004$ to $1.712\pm 0.006$),
and the harder power-law provides a somewhat better fit (see the $\chi^{2}$ values
in Table~\ref{tab:parameters1}).

\begin{figure*}
\plotone{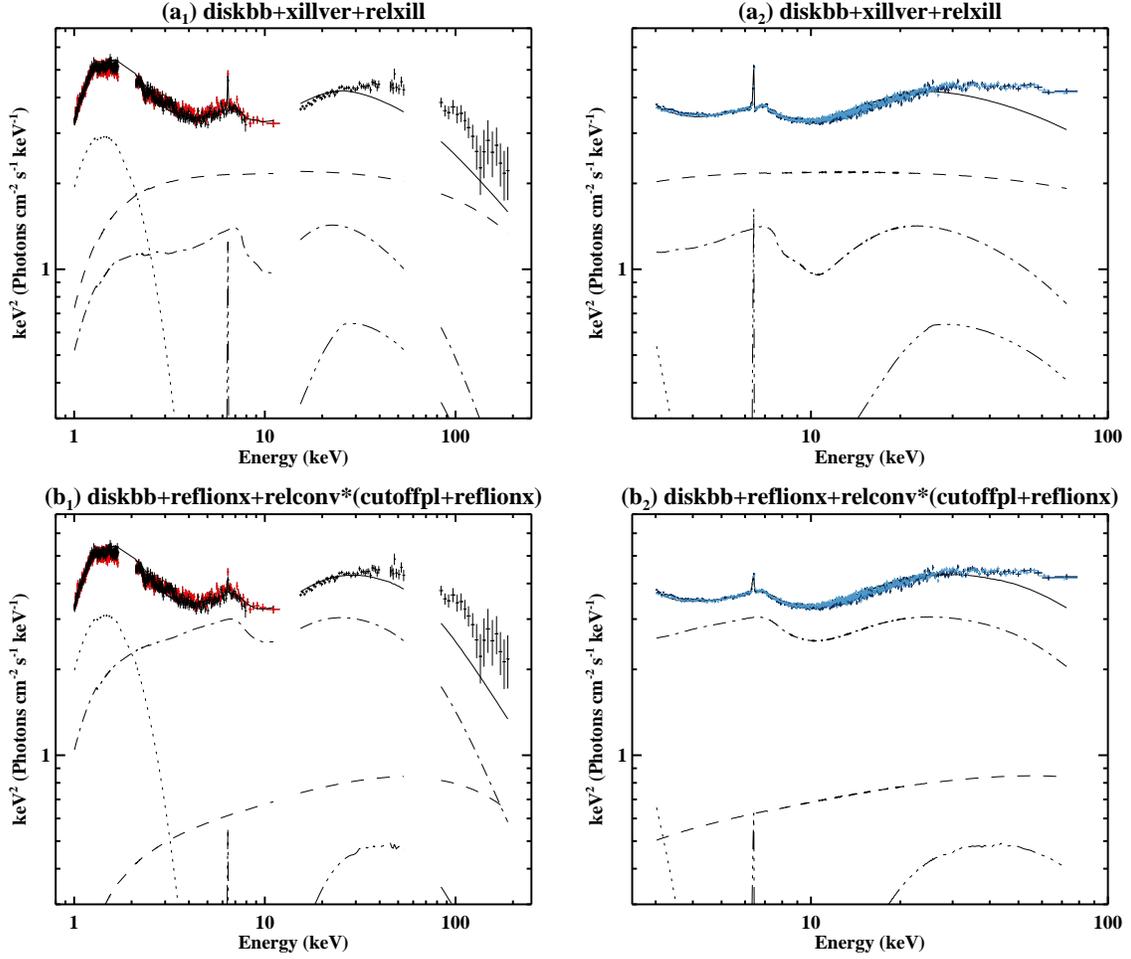}
\caption{\small Unfolded {\em Suzaku} (left panels) and {\em NuSTAR} (right panels)
spectra fitted with two models with the iron abundance parameter fixed to 
$A_{\rm Fe} = 1$.  The components use the same line styles as in Figure~\ref{fig:spectra}, 
but the reflection hump for the {\ttfamily reflionx} model ({\em triple-dot-dashed line})
is visible because the the lower iron abundance forces the overall normalization of this
component to increase.  Although the only change in the model from the spectra shown in
Figure~\ref{fig:spectra} is the factor of $\sim$10 lower $A_{\rm Fe}$, the differences
between the data and the model are largest at the high energy end of the
spectrum.\label{fig:spectra_afe1}}
\end{figure*}

To improve our understanding of why the models require such high iron abundances,
we fixed $A_{\rm Fe}$ to 1.0 (solar abundances) and refit the spectrum.  In both the
{\ttfamily relxill}-based and {\ttfamily reflionx}-based models, the fit is very
poor and, perhaps surprisingly, the largest differences between the models and
the data are at the high energy end rather than in the iron line region (see
Figure~\ref{fig:spectra_afe1}).  In these models, we fixed the cross-normalization
constant for GSO to the values of $C_{\rm GSO}$ given in Table~\ref{tab:parameters1}
because otherwise $C_{\rm GSO}$ would rise to unreasonably high levels.  While both
models demonstrate that low density models with solar abundances cannot fit both
the iron line and the high energy part of the spectrum, the details of why the two
models fail is different.  For the {\ttfamily relxill}-based model, changing to
$A_{\rm Fe} = 1$ causes a drop in the reflection component's soft X-ray emission,
leading to a softening of the power-law index to $\Gamma = 1.963\pm 0.002$, which
causes the model to drop well below the data at high energies.  For the
{\ttfamily reflionx}-based model, the Fe absorption edge is not as deep as for
{\ttfamily relxill}, and the reflection component increases to fit the Fe edge in
the data.  However, the reflection hump turns over at an energy that is much too
low, leaving a large excess at high energies.

Comparing the spectra with $A_{\rm Fe}\sim 10$ (Figure~\ref{fig:spectra}) to
those with $A_{\rm Fe} = 1$ (Figure~\ref{fig:spectra_afe1}), it may be somewhat
surprising that the apparent difference in the iron line strength is not larger.  
However, while increasing $A_{\rm Fe}$ does cause the emission line flux to 
rise, the absorption also increases, and this can be most easily seen by the
fact that the iron absorption edge is much deeper for the spectra shown in
Figure~\ref{fig:spectra}.  The continuum absorption also increases, and we have
checked that if we increase $A_{\rm Fe}$ in the {\ttfamily relxill} model while
leaving all other parameters fixed, the model flux drops above and below the 
iron line, leading to an increase in the iron line equivalent width (EW).  
The details of how the EW depends on $A_{\rm Fe}$ as well as the ionization 
parameter {$\xi$} have been well-documented for the {\ttfamily xillver} model 
\citep{gkm11}.  The EW increases with $A_{\rm Fe}$ but decreases as the disk
becomes more ionized.

We produced a high density model called {\ttfamily reflionx\_hd}, which is an
XSPEC table model using the same code described in \cite{rf07}.  The free
parameters in the model are the power-law photon index ($\Gamma$), the ionization
parameter ($\xi$), and the density ($n_{e}$).  Models are calculated for 10 values
of $\Gamma$ between 1.4 and 2.3, 10 values of $\xi$ between 100 and
10000\,erg\,cm\,s$^{-1}$, and 16 values of $n_{e}$ between $10^{15}$ and
$10^{22}$\,cm$^{-3}$.  An important difference between {\ttfamily reflionx\_hd} and
the model described in \cite{rf07} is that the \cite{rf07} model includes blackbody
irradiation that is an additional heat source for the disk, but we do not include
this source for {\ttfamily reflionx\_hd}.  We emphasize that $A_{\rm Fe} = 1$ for
this model, so the high-density effects can be seen by comparing the results to
the fits shown in Figures~\ref{fig:spectra_afe1}b$_{1}$ and \ref{fig:spectra_afe1}b$_{2}$.

\begin{figure*}
\plotone{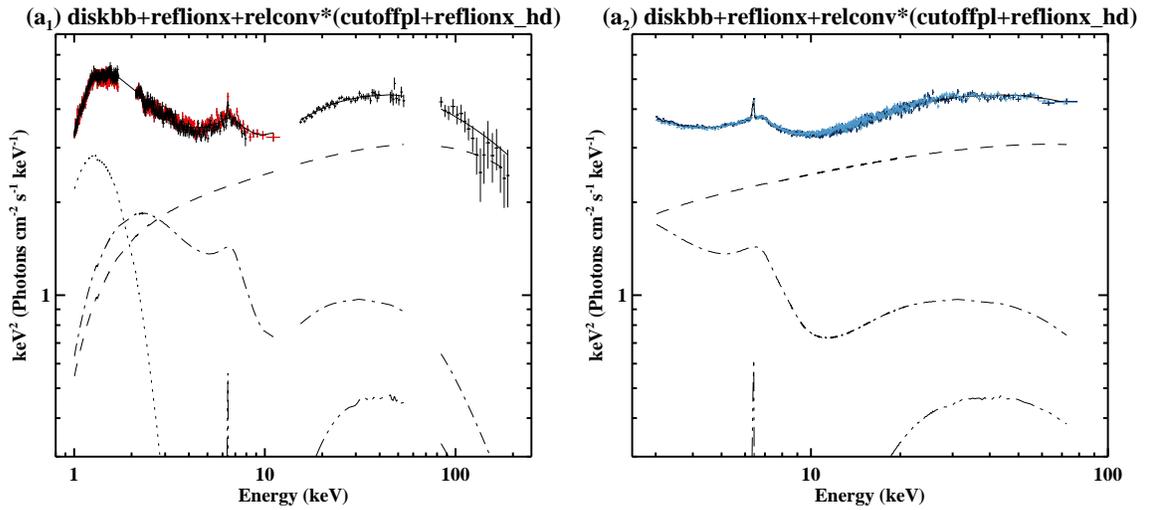}
\caption{\small Unfolded {\em Suzaku} (left panels) and {\em NuSTAR} (right panels)
spectra fitted with a model that has solar abundances ($A_{\rm Fe} = 1$), but with
an electron density of $n_{e} = 4\times 10^{20}$\,cm$^{-3}$ (compared to
$10^{15}$\,cm$^{-3}$ for Figures~\ref{fig:spectra} and \ref{fig:spectra_afe1}).  The
reflection component ({\em dash-dotted line}) has a larger soft excess due to the
higher electron density.  The residuals are shown in Figure~\ref{fig:ratio}d$_{1}$
and \ref{fig:ratio}d$_{2}$.\label{fig:spectra_hd}}
\end{figure*}

Using {\ttfamily reflionx\_hd} provides a very large improvement in the fit.  The
parameters for the fit are provided in Table~\ref{tab:parameters2}, and the quality
of the fit is the best that we have obtained thus far ($\chi^{2}/\nu$ = 3020/2752).
This is also demonstrated in Figures~\ref{fig:ratio}d$_{1}$ and \ref{fig:ratio}d$_{2}$,
which shows that this model removes the positive residuals at the high-energy end of
the {\em NuSTAR} bandpass as well as the dip near 7\,keV.  The density is
well-constrained to be $n_{e} = (3.98^{+0.12}_{-0.25})\times 10^{20}$\,cm$^{-3}$
(see Table~\ref{tab:parameters2}), and the components for this model are shown in
Figure~\ref{fig:spectra_hd}.  The additional soft X-ray emission caused by the higher
density is apparent.  The soft excess peaks at an energy above the 
{\ttfamily diskbb} temperature, and we note that a change in $kT_{\rm in}$ does 
not lead to a model that fits both the thermal disk component and the soft excess.
Although we originally left $q_{\rm in}$ and $R_{\rm break}$ as free parameters in the 
model, they were not well-constrained, and we fixed $q_{\rm in}$ to be equal to 
$q_{\rm out}$ (so they were both fixed to 3).  At least in part, the emissivity 
parameters became poorly constrained because the inner disk radius parameter
increased (ultimately, we obtained $R_{\rm in} = 7.3^{+4.6}_{-1.9}$\,$R_{\rm ISCO}$), but
the emissivity is assumed to drop to zero inside $R_{\rm in}$.  

In order to check on whether the source geometry has a significant impact on the results,
we also fit the spectrum with a lamppost geometry by changing the {\ttfamily relconv}
convolution model to {\ttfamily relconv\_lp}, while continuing to use the high density
reflection model ({\ttfamily reflionx\_hd}).  With the lamppost, the quality of the fit
is equivalent ($\chi^{2}/\nu$ = 3016/2751), and most parameters, including the density,
show very little or no changes.  The lamppost height is
$h = 20.0^{+15.9}_{-1.0}$\,$R_{\rm ISCO}$, and the inner disk radius is
$R_{\rm in} = 2.9^{+3.5}_{-0.8}$\,$R_{\rm ISCO}$ (see Table~\ref{tab:parameters2}).
Although this still suggests a slightly truncated disk, changing to the lamppost
model causes the 90\% confidence lower bound on $R_{\rm in}$ to move from 
5.4\,$R_{\rm ISCO}$ to 2.1\,$R_{\rm ISCO}$ (considering errors).

\section{Discussion}

We have focused on a broadband spectrum of Cyg~X-1 in the intermediate 
state with relatively high statistical quality to investigate current
uncertainties related to accreting black holes at intermediate luminosities.
These include questions about the disk geometry, including understanding
when disks become truncated and whether they are warped, as well as about
the supersolar iron abundances that have been inferred from reflection
fits.  To place the intermediate state observation in context of the
previous reports of {\em NuSTAR} and {\em Suzaku} observations of Cyg~X-1,
we compile some key parameters in Table~\ref{tab:compare}.  As expected,
the values of the disk-blackbody temperature and $\Gamma$ are intermediate
between the hard state value from \cite{basak17} and the soft state values
from \cite{tomsick14} and \cite{walton16}.  The 0.5--100\,keV luminosity 
for the intermediate state observation is $1.8\times 10^{37}$\,erg\,s$^{-1}$, 
which corresponds to $L/L_{\rm Edd} = 0.94$\% for a 14.8\Msun~BH.  In 
comparison, the hard and soft state values are 0.62\% and 1.72\%, 
respectively.  

\begin{figure*}
\plotone{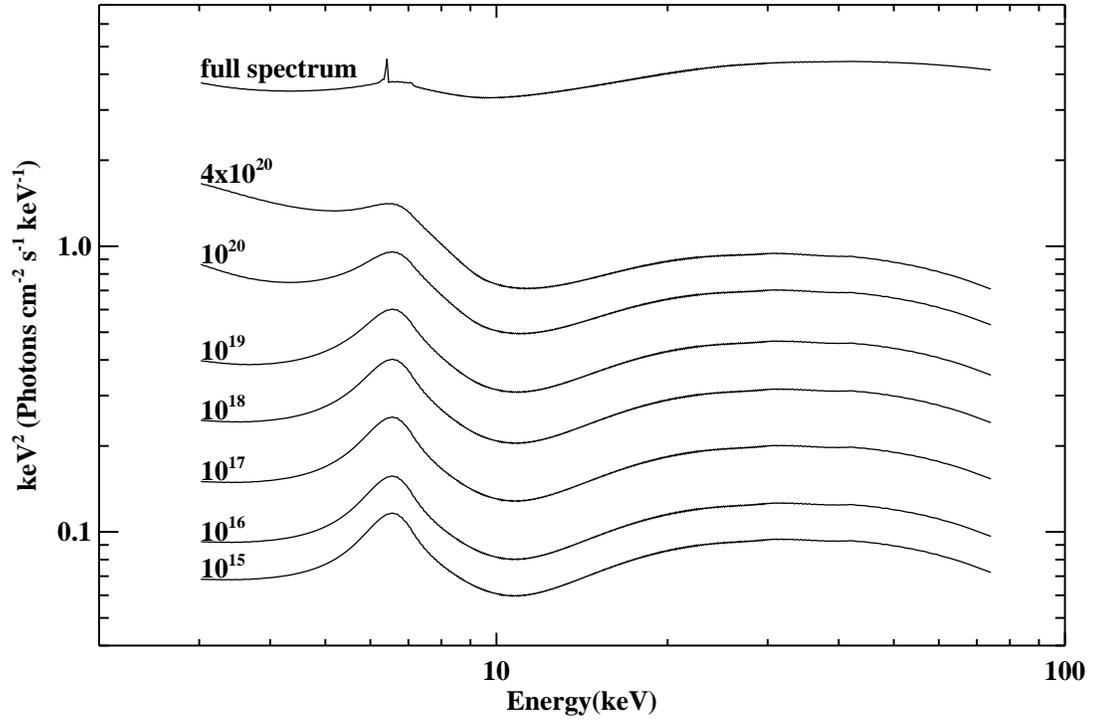}
\caption{\small Relativistic reflection models in the {\em NuSTAR} bandpass using
the {\ttfamily reflionx\_hd} model with values of the electron density ranging 
from $n_{e} = 10^{15}$ to $4\times 10^{20}$\,cm$^{-3}$.  The highest density model
is a component in the full spectrum.\label{fig:model_hd}}
\end{figure*}

As described above, previous observations of Cyg~X-1 have shown supersolar
abundances, but the intermediate state observation marks the first time
that values as high as 10 times solar have been seen in Cyg~X-1, prompting
us to look for other explanations.  A possible explanation that we study 
in some detail here is that the high $A_{\rm Fe}$ is related to high-density 
effects, and our results show that a high-density model allows $A_{\rm Fe}$
to drop to solar abundances while actually improving the fit to the spectrum.
The improvement in the fit occurs because the higher density produces 
extra soft X-ray emission, allowing for a harder direct power-law, which 
provides a better match to the high-energy part of the spectrum.
Figure~\ref{fig:model_hd} shows that the soft excess caused by the increase
in free-free absorption only begins to be significant in the {\em NuSTAR} 
bandpass when densities of $n_{e}\gsim 10^{20}$\,cm$^{-3}$ are reached.  
While \cite{garcia16} describe this effect, that work also emphasizes that
the atomic physics that goes into the reflection models is only known to be
accurate up to densities of $10^{19}$\,cm$^{-3}$.  Thus, a caveat to our
results is that more accurate determinations of quantities such as the rates
of atomic transitions could have an impact on the high density reflection
models.

While the high-density explanation for Cyg~X-1 makes sense because estimates
show that BH binary disks should have densities that are much higher than
$10^{15}$\,cm$^{-3}$ \citep{sz94,garcia16}, it is not a unique explanation
for the high $A_{\rm Fe}$ values.  We have also produced a model with
$n_{e} = 10^{15}$\,cm$^{-3}$ but with blackbody irradiation of the disk, and
this also provides a good fit to the spectrum with $A_{\rm Fe} = 1$.  However,
the good fit is obtained for a blackbody temperature of 0.7\,keV, which is
hotter than the inner disk temperature of $kT_{\rm in} = 0.3$--0.4\,keV that
we measure directly.  Thus, this model is not physically self-consistent,
but it does show another way that the addition of a soft excess component
can provide a good fit to the spectrum with solar iron abundances.  

In other papers on fits to Cyg~X-1 spectra in the hard or intermediate state,
models with two Compton continuum components have been used.  \cite{yamada13b}
and \cite{basak17} both fit broadband Cyg~X-1 spectra with hard and soft
Comptonization components.  We have applied such models to our spectrum, and
they also provide good fits with solar iron abundances.  However, they have many
more free parameters than our {\ttfamily reflionx\_hd} model, and the physical
picture is more complicated as it is unclear that coronae with temperatures in
the 1--5\,keV range \citep[e.g., as found by][]{basak17} exist.  There are other 
physical scenarios which might produce multiple high-energy components such as 
a disk corona and a jet, and these have previously been applied to Cyg~X-1 
spectra \citep[e.g.,][]{nowak11}, but they are also more complicated and have 
many more free parameters than the {\ttfamily reflionx\_hd} model.

Regardless of how the extra emission below the iron line is produced, it
has an impact on the constraints on the inner radius of the disk because it
is the gravitationally redshifted wing of the line that provides the power to
measure $R_{\rm in}$.  This can be seen in our spectral fits since the
high-$A_{\rm Fe}$ fits give inner radii of $R_{\rm in} = 1.3$\,$R_{\rm ISCO}$
(see Table~\ref{tab:parameters1}), which are very close to the ISCO, while
the high-density fits show larger inner radii.  Table~\ref{tab:parameters2}
lists values of $R_{\rm in} = 7.3^{+4.6}_{-1.9}$\,$R_{\rm ISCO}$ for the broken
power-law emissivity and $R_{\rm in} = 2.9^{+3.5}_{-0.8}$\,$R_{\rm ISCO}$ for
the lamppost model.  

The high-density models also affect the values obtained for the inner disk
inclination.  While we have previously found values that are significantly
larger than the binary inclination \citep{tomsick14,walton16}, and our
high-$A_{\rm Fe}$ fits in this work also give high inclinations, the
high-density fits give $18^{+4}_{-5}$$^{\circ}$ for the broken power-law
emissivity and a 90\% confidence upper limit of $<$$20^{\circ}$ for the
lamppost model.  Using the lamppost model, we performed an additional fit to
the data with the inner disk inclination fixed to the binary inclination
of $27^{\circ}$.  While this does degrade the quality of the fit somewhat
($\chi^{2}/\nu = 3036/2752$ compared to 3016/2751), the residuals are still
relatively small.

As the density of AGN disks is estimated to be close to the standard value
of $n_{e} = 10^{15}$\,cm$^{-3}$, we would expect the AGN iron abundances to
be closer to the true values.  However, there are several examples of
supersolar iron abundances obtained by fitting their spectra with reflection
models.  The AGN abundances range from moderately supersolar values, such
as $A_{\rm Fe} = 2$--4 for NGC~3783 \citep{reynolds12} and $A_{\rm Fe}\sim 3$
for NGC~1365 \citep{risaliti13} to values of $A_{\rm Fe} = 9$ for
1H~0707--495 \citep{fabian09} or even higher for IRAS~13224--3809
\citep{fabian13}.  In at least some cases, the supersolar abundances may
be real.  Looking at optical line properties of quasars, \cite{wang12}
show evidence for abundances up to 7 times solar, and \cite{reynolds12}
hypothesize that radiative levitation can enhance iron abundances.  We
discuss this possibility in Section 3 and argue that this effect would
cause rapid changes in $A_{\rm Fe}$ for X-ray binaries.  However, this
would probably not be a problem for AGN due to the longer time scales.

In addition to the caveats discussed above about the atomic physics being
uncertain above $10^{19}$\,cm$^{-3}$, about the high-density model not
being a unique solution, and about the high iron abundances in AGN, it is
also important to point out that we have only looked at the high-density
model for the case of solar abundances.  While we can rule out large changes
in $A_{\rm Fe}$, implying that abundances of 10 times solar are unreasonable
for Cyg~X-1, we cannot rule out supersolar abundances at lower levels.
Thus, we caution against over-interpretation of the inner radius and
inclination values.  In the near future, it is important to improve the
atomic physics in the high-density models and also to make reflection models
which consider a range of $A_{\rm Fe}$ values.  With such models, it would be
worthwhile to expand the analysis to more Cyg~X-1 spectra as well as other
sources where fit parameters suggest the possibility of supersolar abundances.

\acknowledgments

This work made use of data from the {\it NuSTAR} mission, a project led by the
California Institute of Technology, managed by the Jet Propulsion Laboratory,
and funded by the National Aeronautics and Space Administration. We thank the
{\it NuSTAR} Operations, Software and  Calibration teams for support with the
execution and analysis of these observations.  This research has made use of
the {\it NuSTAR}  Data Analysis Software (NuSTARDAS) jointly developed by the
ASI Science Data Center (ASDC, Italy) and the California Institute of Technology
(USA).  This research has made use of data obtained from the {\em Suzaku} 
satellite, a collaborative mission between the space agencies of Japan (JAXA) 
and the USA (NASA).  JAT acknowledges partial support from {\em NuSTAR} Guest 
Observer grant NNX15AV23G.  ACF acknowledges support from ERC Advanced Grant
340442.  DJW acknowledges support from STFC in the form of an Ernest Rutherford
fellowship.  JAT thanks K.~Hamaguchi for help with the XIS data 
reduction and T.~Dauser, K.~Koljonen, and A.~Zoghbi for useful discussions.
This research has made use of the {\em MAXI} data provided by RIKEN, JAXA, and
the {\em MAXI} team.




\begin{table*}
\caption{Observing Log and Exposure Times\label{tab:obs}}
\begin{minipage}{\linewidth}
\begin{center}
\begin{tabular}{cccccc} \hline \hline
        &            &       & Start Time (UT) & End Time (UT) & Livetime\\
Mission & Instrument & ObsID & (in 2015)       & (in 2015)     & (s)\\
\hline\hline
{\em NuSTAR} & FPMA     & 30101022002 & May 27, 17.3 h  & May 28, 11.8 h & 19,860\\
{\em NuSTAR} & FPMB     &   ''        &       ''        &           ''   & 20,500\\
{\em Suzaku} & XIS0     & 410018010   & May 27, 17.2 h  & May 28, 15.6 h & 1,098\\
{\em Suzaku} & XIS1     &      ''     & May 27, 16.7 h  & May 28, 18.8 h & 4,991\\
{\em Suzaku} & XIS3     &      ''     & May 27, 16.7 h  & May 28, 15.1 h & 2,913\\
{\em Suzaku} & HXD/PIN  &      ''     & May 27, 17.4 h  & May 28, 18.1 h & 18,620\\
{\em Suzaku} & HXD/GSO  &      ''     &       ''        &          ''    & 18,620\\ \hline
\end{tabular}
\end{center}
\end{minipage}
\end{table*}

\begin{table*}
\caption{Fit Parameters for Reflection Models with Free Iron Abundnace\label{tab:parameters1}}
\begin{minipage}{\linewidth}
\begin{center}
\begin{tabular}{cccc} \hline \hline
Parameter & Unit/Description & Value\footnote{With 90\% confidence errors.} & Value$^{a}$\\ 
          &                  & ({\ttfamily relxill}-based) & ({\ttfamily reflionx}-based)\\ \hline
\multicolumn{4}{l}{Interstellar Absorption}\\
$N_{\rm H}$ & $10^{21}$\,cm$^{-2}$ & 6.0\footnote{Fixed.} & 6.0$^{b}$\\ \hline
\multicolumn{4}{l}{Disk-blackbody}\\
$kT_{\rm in}$ & keV & $0.378\pm 0.002$ & $0.376\pm 0.002$\\
$N_{\rm DBB}$ & Normalization & $44,600\pm 1100$ & $46,600^{+1500}_{-1400}$\\ \hline
\multicolumn{4}{l}{Direct component plus partially ionized/relativistic reflection}\\
$\Gamma$ & Photon Index & $1.826\pm 0.004$ & $1.712\pm 0.006$\\
$E_{\rm cut}$ & keV & 300$^{b}$ & 300$^{b}$\\
$\log{\xi}$ & $\xi$ in erg\,cm\,s$^{-1}$ & $3.374\pm 0.013$ & $3.588^{+0.044}_{-0.035}$\\
$A_{\rm Fe}$ & Abundance relative to solar & $>$9.96 & $10.6^{+1.6}_{-0.9}$\\
$\Omega/2\pi$ & Covering fraction & $0.87\pm 0.04$ & ---\\
$q_{\rm in}$ & Emissivity Index & $9.1^{+0.5}_{-0.4}$ & $9.3\pm 0.3$\\
$q_{\rm out}$ & Emissivity Index & 3$^{b}$ & 3$^{b}$\\
$R_{\rm break}$ & Index Break Radius ($R_{\rm ISCO}$) & $2.24^{+0.07}_{-0.22}$ & $2.41^{+0.06}_{-0.22}$\\
$R_{\rm in}$ & Disk Inner Radius ($R_{\rm ISCO}$) & $1.33^{+0.03}_{-0.06}$ & $1.26^{+0.03}_{-0.02}$\\
$R_{\rm out}$ & Outer Radius for Reflection ($R_{\rm g}$) & 400$^{b}$ & 400$^{b}$\\
$a_{*}$ & Black Hole Spin & $>$0.987 & $>$0.988\\
$i$ & Inclination ($^{\circ}$) & $37.5\pm 0.7$ & $35.6\pm 1.0$\\
$N_{\rm relxill}$ & Normalization & $(3.74^{+0.04}_{-0.05})\times 10^{-2}$ & ---\\ 
$N_{\rm reflionx}$ & Normalization & --- & $(1.88^{+0.13}_{-0.15})\times 10^{-4}$\\
$N_{\rm cutoffpl}$ & Normalization & --- & $26.2^{+0.7}_{-1.3}$\\ \hline
\multicolumn{4}{l}{Neutral reflection\footnote{The model assumes that there is a single direct component that produces both the partially ionized/relativistic and neutral reflection components.  Also, $i$ and $A_{\rm Fe}$ are assumed to be the same for the partially ionized/relativistic and neutral reflection components.}}\\
$\log{\xi}$ & $\xi$ in erg\,cm\,s$^{-1}$ & 0.0$^{b}$ & 1.0$^{b}$\\
$N_{\rm xillver}$ & Normalization & $(1.77^{+0.22}_{-0.25})\times 10^{-3}$ & ---\\
$N_{\rm reflionx}$ & Normalization & --- & $(3.5\pm 0.4)\times 10^{-4}$\\ \hline
\multicolumn{4}{l}{Cross-Normalization Constants (relative to XIS0+XIS3)}\\
$C_{\rm XIS1}$ & -- & $0.794\pm 0.003$ & $0.794\pm 0.003$\\
$C_{\rm FPMA}$ & -- & $1.005\pm 0.004$ & $1.002\pm 0.004$\\
$C_{\rm FPMB}$ & -- & $1.007\pm 0.004$ & $1.003\pm 0.004$\\
$C_{\rm PIN}$ & -- & $1.247\pm 0.007$ & $1.244\pm 0.007$\\
$C_{\rm GSO}$ & -- & $1.13\pm 0.03$   & $1.15\pm 0.03$\\ \hline
\multicolumn{4}{l}{Goodness of fit}\\
$\chi^{2}/\nu$ & -- & 3566/2750 & 3360/2750\\ \hline
\end{tabular}
\end{center}
\end{minipage}
\end{table*}

\begin{table*}
\caption{Fit Parameters for High Density Reflection Model\label{tab:parameters2}}
\begin{minipage}{\linewidth}
\begin{center}
\begin{tabular}{cccc} \hline \hline
Parameter & Unit/Description & Value\footnote{With 90\% confidence errors.} & Value$^{a}$\\
          &                  & (power-law emissivity) & (lamppost)\\ \hline
\multicolumn{4}{l}{Interstellar Absorption}\\
$N_{\rm H}$ & $10^{21}$\,cm$^{-2}$ & 6.0\footnote{Fixed.} & 6.0$^{b}$\\ \hline
\multicolumn{4}{l}{Disk-blackbody}\\
$kT_{\rm in}$ & keV & $0.317\pm 0.002$ & $0.317\pm 0.002$\\
$N_{\rm DBB}$ & Normalization & $100,500^{+3700}_{-3500}$ & $100,400^{+2500}_{-3400}$\\ \hline
\multicolumn{4}{l}{Direct component plus partially ionized/relativistic reflection}\\
$\Gamma$ & Photon Index & $1.779\pm 0.007$ & $1.777^{+0.009}_{-0.006}$\\
$E_{\rm cut}$ & keV & 300$^{b}$ & 300$^{b}$\\
$\log{\xi}$ & $\xi$ in erg\,cm\,s$^{-1}$ & $3.302^{+0.011}_{-0.006}$ & $3.303^{+0.014}_{-0.002}$\\
$A_{\rm Fe}$ & Abundance relative to solar & 1.0\footnote{The {\ttfamily reflionx\_hd} model is for solar abundances.} & 1.0$^{c}$\\
$q_{\rm in}$ & Emissivity Index & 3$^{b}$ & ---\\
$q_{\rm out}$ & Emissivity Index & 3$^{b}$ & ---\\
$h$ & Lamppost height ($R_{\rm ISCO}$) & --- & $20.0^{+15.9}_{-1.0}$\\
$R_{\rm in}$ & Disk Inner Radius ($R_{\rm ISCO}$) & $7.3^{+4.6}_{-1.9}$ & $2.9^{+3.5}_{-0.8}$\\
$R_{\rm out}$ & Outer Radius for Reflection ($R_{\rm g}$) & 400$^{b}$ & 400$^{b}$\\
$a_{*}$ & Black Hole Spin &  $>$0.93 & $0.949^{+0.013}_{-0.019}$\\
$i$ & Inclination ($^{\circ}$) & $18^{+4}_{-5}$ & $<$20\\
$n_{e}$ & Density in $10^{20}$\,cm$^{-3}$ & $3.98^{+0.12}_{-0.25}$ & $3.98^{+0.13}_{-0.33}$\\
$N_{\rm reflionx}$ & Normalization & $97.9^{+2.9}_{-1.5}$ & $97.8^{+4.0}_{-1.5}$\\
$N_{\rm cutoffpl}$ & Normalization & $24.8\pm 0.4$ & $24.8^{+0.4}_{-0.3}$\\ \hline
\multicolumn{4}{l}{Neutral reflection\footnote{The model assumes that there is a single direct component that produces both the partially ionized/relativistic and neutral reflection components.  Also, $i$ and $A_{\rm Fe}$ are assumed to be the same for the partially ionized/relativistic and neutral reflection components.}}\\
$\log{\xi}$ & $\xi$ in erg\,cm\,s$^{-1}$ & 1.0$^{b}$ & 1.0$^{b}$\\
$N_{\rm reflionx}$ & Normalization & $(8.6\pm 0.8)\times 10^{-4}$ & $(8.3^{+0.2}_{-0.7})\times 10^{-4}$\\ \hline
\multicolumn{4}{l}{Cross-Normalization Constants (relative to XIS0+XIS3)}\\
$C_{\rm XIS1}$ & -- & $0.793\pm 0.003$ & $0.793^{+0.001}_{-0.003}$\\
$C_{\rm FPMA}$ & -- & $1.002\pm 0.004$ & $1.002^{+0.004}_{-0.001}$\\
$C_{\rm FPMB}$ & -- & $1.004\pm 0.004$ & $1.004^{+0.004}_{-0.001}$\\
$C_{\rm PIN}$ & -- & $1.244\pm 0.007$ & $1.244\pm 0.007$\\
$C_{\rm GSO}$ & -- & $1.02\pm 0.03$ & $1.02\pm 0.03$\\ \hline
\multicolumn{4}{l}{Goodness of fit}\\
$\chi^{2}/\nu$ & -- & 3020/2752 & 3016/2751\\ \hline
\end{tabular}
\end{center}
\end{minipage}
\end{table*}

\begin{table*}
\caption{Key Parameters for Cyg~X-1 in different Spectral States\label{tab:compare}}
\begin{minipage}{\linewidth}
\begin{center}
\begin{tabular}{cccccc} \hline \hline
         & \cite{tomsick14}  & \cite{walton16} & This work & This work & \cite{basak17}\\
         & ({\em NuSTAR} and & (4 {\em NuSTAR} & {\ttfamily relxill} & {\ttfamily reflionx\_hd} & ({\em NuSTAR} and\\
         & {\em Suzaku})     & observations)   &                     & (lamppost) & {\em Suzaku})\\ \hline
Spectral State & soft & soft & intermediate & intermediate & hard\\
$kT_{\rm in}$ (keV)  & 0.56 & 0.40--0.47 & $0.378\pm 0.002$ & $0.317\pm 0.002$ & 0.14--0.15\\
$\Gamma$       & 2.6--2.7 & 2.56--2.74 & $1.826\pm 0.004$ & $1.777^{+0.009}_{-0.006}$ & 1.70--1.71\\
$R_{\rm in}/R_{\rm ISCO}$ & 1 & 1 & $1.33^{+0.03}_{-0.06}$ & $2.9^{+3.5}_{-0.8}$ & 13--20\,$R_{g}$\footnote{13--20 gravitational radii, $R_{g}$, corresponds to 13--20\,$R_{\rm ISCO}$ for a maximally rotating BH, 6.5--10\,$R_{\rm ISCO}$ for a BH rotating at $a_{*} = 0.94$, and 2.2--3.3\,$R_{\rm ISCO}$ for a non-rotating BH.}\\
$a_{*}$ & $>$0.83 & 0.93--0.96 & $>$0.987 & $0.949^{+0.013}_{-0.019}$ & ---\\
$i$ ($^{\circ}$) & $>$40 & 38.2--40.8 & $37.5\pm 0.7$ & $<$20 & 24--43\\
$A_{\rm Fe}$ & 1.9--2.9 & 4.0--4.3 & $>$9.96 & 1.0 & 2.2--4.6\\
Flux\footnote{Unabsorbed flux in the 0.5--100\,keV band in erg\,cm$^{-2}$\,s$^{-1}$.} & $8.0\times 10^{-8}$ & ---\footnote{We do not quote fluxes or luminosities for these four observations.  Only {\em NuSTAR} was used in the analysis, and the disk-blackbody component is not constrained well enough to extrapolate down to 0.5\,keV.} & $4.4\times 10^{-8}$ & ---\footnote{The fluxes and luminosities are not significantly different for {\ttfamily reflionx\_hd} compared to {\ttfamily relxill}.} & $2.9\times 10^{-8}$\\
Luminosity\footnote{In erg\,s$^{-1}$ using the 0.5--100\,keV unabsorbed flux and a distance of 1.86\,kpc.} & $3.3\times 10^{37}$ & ---$^{c}$ & $1.8\times 10^{37}$ & ---$^{d}$ & $1.2\times 10^{37}$\\
$L/L_{\rm Edd}$ & 1.72\% & ---$^{c}$ & 0.94\% & ---$^{d}$ & 0.62\%\\ \hline
\end{tabular}
\end{center}
\end{minipage}
\end{table*}

\end{document}